\documentclass[aps,pra,showpacs,twocolumn]{revtex4}
\usepackage{epsfig}
\begin{document}
\title{Quantum partial teleportation as optimal cloning at a distance}
\author{Radim Filip}
\affiliation{Department of Optics, Palack\' y University,\\
17. listopadu 50,  772~07 Olomouc, \\ Czech Republic}
\date{\today}
\begin{abstract}
We propose a feasible scheme of conditional quantum partial teleportation of a qubit as optimal 
asymmetric cloning at a distance. In this scheme, Alice preserves one imperfect clone whereas other clone is teleported to Bob. 
Fidelities of the clones can be simply controlled by an asymmetry in Bell-state measurement. 
The optimality means that tightest inequality for the fidelities in the asymmetric cloning is saturated. 
Further we design a conditional teleportation as symmetric optimal $N\rightarrow N+1$ cloning 
from $N$ Alice's replicas on single distant 
clone. We shortly discussed two feasible experimental implementations, 
first one for teleportation of polarization state of a photon and second one, for teleportation of 
a time-bin qubit.    
\end{abstract}
\maketitle

\section{Introduction}

One from main tasks of quantum information processing is how to optimally distribute an
unknown quantum state $|\Psi\rangle$ of a qubit to another distant qubit. 
A perfect quantum teleportation \cite{teleth} where Alice completely 
transmits unknown qubit state $|\Psi\rangle$ to distant Bob's qubit, 
is a particular example of this task. As a resource, they 
share a pair of qubits in a maximally entangled state which can be distributed apriori 
and then, in advantage, they perform only local operations and classical communication (LOCC) 
in an actual time of the state transmission. After this 
complete quantum teleportation, Alice has no information about input qubit state. 
A conditional version of qubit teleportation of a polarization state of photon was 
experimentally demonstrated using a simple Bell-state analyzer based on balanced beam splitter 
\cite{teleexp}. Recently, 
also conditional long-distance quantum teleportation of a time-bin qubit in telecommunication 
fibers has been realized \cite{Marcikic03}.

In this paper, we extend the conditional teleportation scheme to {\em partial optimal teleportation} 
of single replica of unknown qubit state. The partial teleportation means that Alice preserves an 
imperfect copy $\rho_{S}$ of input state and Bob obtains the other imperfect 
copy $\rho_{S'}$. In the partial teleportation the fidelities of copies 
$F_{S}=\langle\Psi|\rho_{S}|\Psi\rangle$, $F_{S'}=\langle\Psi|\rho_{S'}|\Psi\rangle$
can be controled by an asymmetry in the 
Bell state measurement. 
The optimality in this case says that for a given fidelity of Alice's copy with initial state, 
Bob cannot in principle 
obtain a higher fidelity of his copy. This extension of teleportation 
can be straightforwardly implement into the recent 
conditional teleportation experiments \cite{teleexp}, using the Bell-state analyzer 
beam splitter or fiber coupler with a variable reflectivity. 
Further, we propose a scheme for teleportation from $N$ 
identical replicas of a qubit state to a single distant copy. 

The partial transmission of a quantum state 
is dissimilar to classical information processing because
perfect cloning of an unknown quantum state is impossible \cite{nocloning}. Thus 
the fidelities of the copies after the partial teleportation are limited by the fidelities in 
optimal quantum cloning. Previously, universal symmetric optimal quantum cloners were theoretically 
proposed to {\em locally} distribute an unknown pure quantum state of a qubit to the copies 
\cite{cloningth1,Gisin97,cloningth2,Bruss98,cloningth3} and were also realized experimentally  
\cite{symmclonexp}.
To locally duplicate an unknown quantum state of a qubit with unbalanced fidelities, 
the asymmetric quantum cloners were 
theoretically discussed \cite{asymmclonth}.
The asymmetric $1\rightarrow 2$ optimal cloning produces two copies 
from a single replica of an unknown state and obtained state-independent 
fidelities $F_{S}$ and $F_{S'}$ of the copies saturate cloning inequality \cite{asymmclonth}
\begin{equation}\label{cond}
(1-F_{S})(1-F_{S'})\geq (1/2-(1-F_{S})-(1-F_{S'}))^{2}.
\end{equation}
This inequality sets the tightest no-cloning bound on fidelities 
of $1\rightarrow 2$ cloning device that duplicates an unknown qubit state to 
another qubit with isotropic noise. 
Thus if the equality occurs in (\ref{cond}) then for given fidelity $F_{S}$ one cannot  
obtained a better fidelity $F_{S'}$. Previously experimentally performed symmetric quantum cloning with 
identical fidelities $F_{S,S'}=5/6$ arises as a particular case. 
An enhancement of the fidelities $F_{S'}=5/6$ of a single additional copy 
can be obtained only if we have $N>1$ identical replicas of the input state and implement 
symmetric $N\rightarrow N+1$ cloning \cite{Gisin97,Bruss98}. 
Then a single additional copy of quantum state can be produced 
with fidelity 
\begin{equation}\label{fid1}
F_{N\rightarrow N+1}=\frac{(N+1)^{2}+N}{(N+1)(N+2)},
\end{equation}       
which approaches unity as the number $N$ of replicas increases. 
From the point of view of quantum cloning, 
the schemes proposed below can be also reviewed as a conditional implementation of 
optimal universal {\em quantum cloning at a distance}. 
These proposal can be also view as a new tele-cloning procedure in comparison with Ref.~\cite{telecl}. 

The paper is organized as follows. In the Sec.~II, we design the partial conditional teleportation scheme and 
prove that it represents optimal asymmetric $1\rightarrow 2$ cloning at a distance.
We also discuss local implementation of U-NOT gate, LOCC reversibility of partial teleportation and 
the sequential partial teleportation. Further, in Sec.~III the partial conditional 
symmetric $N\rightarrow N+1$ teleportation is described and it is proved that it produces $N+1$ 
copies with optimal fidelities. Simultaneously, this scheme locally 
realizes optimal U-NOT gate for $N$ multiple replicas of input state. In the last Sec.~IV 
experimental implementations of these schemes for the polarization and time-bin qubits are shortly
discussed.  

\section{Optimal $1\rightarrow 2$ asymmetric cloning at a distance}

\begin{figure}
\centerline{\psfig{width=8.0cm,angle=0,file=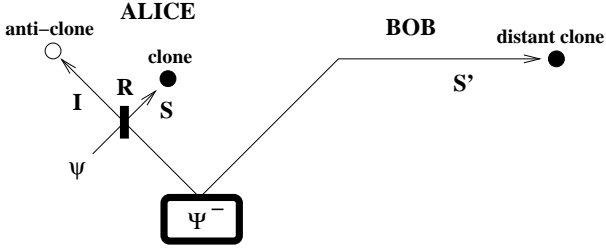}}
\caption{Scheme of conditional partial teleportation as optimal asymmetric $1\rightarrow 2$ 
cloning.}
\end{figure}

A schematic setup for partial conditional teleportation of a qubit is depicted in Fig.~1. In fact, this 
is a feasible modification of the previous experiment on teleportation of a 
polarization state of photon \cite{teleexp}. 
It is based on conditional and partial Bell-state measurement which can be simply implemented by 
an unbalanced beam splitter with a variable reflectivity $R$, 
$0\leq R\leq 1/2$. After mixing two photons $S,I$ on beam splitter we
restrict our teleportation only to such cases when both photons leave the beam splitter separately. 
Then we may effectively describe the unbalanced beam splitter by the following transformation:  
\begin{eqnarray}\label{unBS}
|\Psi\Psi\rangle_{SI} &\rightarrow & (T-R)|\Psi\Psi\rangle_{SI},\nonumber\\
|\Psi\Psi_{\bot}\rangle_{SI}&\rightarrow & T|\Psi\Psi_{\bot}\rangle_{SI}-R|\Psi_{\bot}\Psi
\rangle_{SI},
\end{eqnarray}
which corresponds to the projection 
\begin{equation}\label{proj}
\Pi^{-}_{SI}(R)=\left((1-2R)1_{S}\otimes 1_{I}+2R|\Psi_{-}\rangle_{SI}\langle\Psi_{-}|\right)
\end{equation}
on input polarization state of two photons. 
Assuming that entangled state $|\Psi_{-}\rangle_{IS'}=\frac{1}{\sqrt{2}}(|VH\rangle_{IS'}-
|HV\rangle_{IS'})=\frac{1}{\sqrt{2}}(|\Psi\Psi_{\bot}\rangle_{IS'}-
|\Psi_{\bot}\Psi\rangle_{IS'})$ is shared 
by Alice and Bob, we can prove that Alice can conditionally perform partial 
teleportation of an unknown qubit state $|\Psi\rangle_{S}$ to Bob.
Performing projective measurement $\Pi^{-}_{SI}(R)\otimes 1_{S'}$ on a state of total system 
$|\Psi_{S}\rangle|\Psi_{-}\rangle_{IS'}$
we obtain the following local states of clones $S,S'$ and anti-clone $I$ 
\begin{eqnarray}
\rho_{S,S'}(R)&=&F_{S,S'}(R)|\Psi\rangle\langle\Psi|+
(1-F_{S,S'}(R))|\Psi_{\bot}\rangle\langle\Psi_{\bot}|,\nonumber\\
\rho_{I}(R)&=&(1-F_{I}(R))|\Psi\rangle\langle\Psi|+F_{I}(R)|\Psi_{\bot}\rangle\langle\Psi_{\bot}|\nonumber\\
\end{eqnarray}
with the following fidelities:  
\begin{eqnarray}\label{fide}
F_{S}(R)&=&\frac{1}{2P(R)}\left( (1-2R)^2+(1-R)^2\right),\nonumber\\
F_{S'}(R)&=&\frac{1}{2P(R)}\left( R^2+(1-R)^2\right),\,
F_{I}(R)=\frac{(1-R)^2}{2P(R)},\nonumber\\
\end{eqnarray} 
where $P(R)=1-3R+3R^2$. It can be proved that the fidelities $F_{S}$ 
and $F_{S'}$ saturate the inequality (\ref{cond}) and therefore 
the distribution of input state between clone $S$ and distant clone $S'$ 
is optimal. The symmetric distribution can be obtained for the reflectivity $R=1/3$. 
In this case we also obtain optimal U-NOT gate with fidelity $F_{UNOT}=2/3$ if we take the anticlone $I$ 
as output of the U-NOT. This U-NOT optimally approximates a transformation 
$|\Psi\rangle\rightarrow |\Psi_{\bot}\rangle$ \cite{U-NOT}, only by mixing the input state 
with the random mixed state on unbalanced beam splitter with $R=1/3$. 

Now we show that we can probabilistically 
transform any asymmetric cloner with $R,T\not= 0$ to complete conditional 
teleportation with unit fidelity only by local measurements on Alice's clone and ancilla, 
classical communication with Bob and state filtration on Bob's qubit. 
Assuming input state $|\Psi\rangle_{S}=\alpha|V\rangle_{S}+\beta|H\rangle_{S}$, the state after 
projection (\ref{proj}) can be expanded in the following way: 
\begin{eqnarray}
\alpha (1-2R)|VVH\rangle_{SIS'}-\beta (1-2R) |HHV\rangle_{SIS'} -\nonumber\\
\alpha (1-R)|VHV\rangle_{SIS'} +\alpha R |HVV\rangle_{SIS'} +\nonumber\\
\beta (1-R)|HVH\rangle_{SIS'} 
-\beta R|VHH\rangle_{SIS'}. 
\end{eqnarray}  
Generalizing an idea of the state restoration from Ref.~(\cite{Bruss01}), 
we can measure polarization in basis 
$|V\rangle,|H\rangle$ on Alice's clone and ancilla and the results send to Bob. 
The measurement can be experimentally implemented using polarization beam splitter followed 
on both outputs by single photon detectors which is, in fact, an asymmetric version 
of Bell state measurement in teleportation experiment \cite{Bell}. 
If we select only such results when the orthogonal polarizations 
$|V\rangle_{S}|H\rangle_{I}$ ($|H\rangle_{S}|V\rangle_{I}$) are detected then 
the Bob's state changes to new one, proportional to 
$\alpha(1-R)|V\rangle_{S'}-\beta R |H\rangle_{S'}$ ($\alpha R|V\rangle_{S'}+\beta (1-R) |H\rangle_{S'}$).
These states can be conditionally transform to initial state $|\Psi\rangle_{S}$ 
by the local filtering $R|V\rangle_{S'}\langle V|-(1-R)
|H\rangle_{S'}\langle H|$ ($R|H\rangle_{S'}\langle H|+(1-R)
|V\rangle_{S'}\langle V|$).  On the other hand, Bob can help Alice to 
conditionally restore initial state on her clone. 
Bob has to perform measurement in basis $|V\rangle,|H\rangle$ 
on qubit $S'$ and Alice the same measurement on qubit $I$. 
If the detected state is $|H\rangle_{I}|V\rangle_{S'}$ ($|V\rangle_{I}|H\rangle_{S'}$) then a state of 
the Alice clone is converted to state proportional to 
$\alpha (1-R)|V\rangle_{S}-\beta (1-2R)|H\rangle_{S}$ 
($\alpha (1-2R)|V\rangle_{S}+\beta (1-R)|H\rangle_{S}$) which is equal 
to initial state after conditional state projection 
$(1-2R)|V\rangle_{S}\langle V|-(1-R)|H\rangle_{S}\langle H|$ ($(1-2R)|H\rangle_{S}\langle H|+(1-R)
|V\rangle_{S}\langle V|$). 
Thus we can at least conditionally prove in feasible experiment that 
asymmetric cloning procedure is conditionally LOCC reversible. 

\begin{figure}
\centerline{\psfig{width=8.0cm,angle=0,file=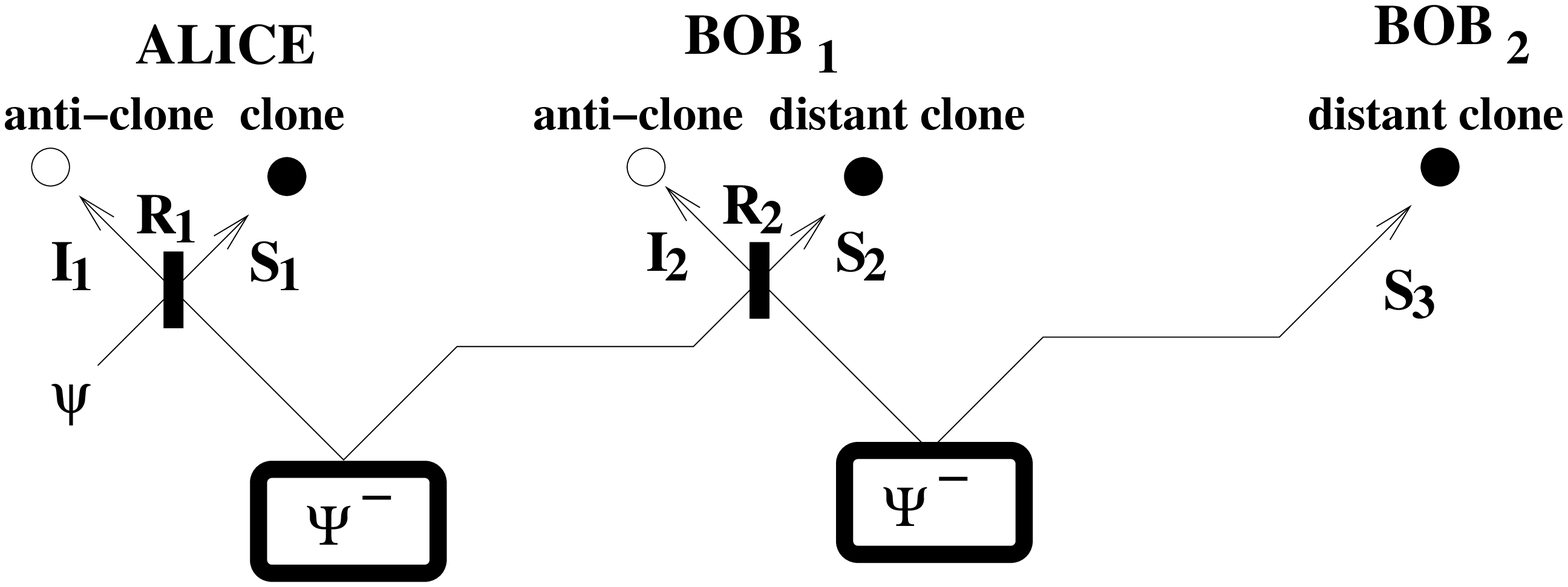}}
\caption{Scheme of sequential conditional partial teleportation.}
\end{figure}

We shortly discuss a sequence of two partial teleportations in which Alice can 
conditionally symmetricly distribute an unknown quantum state 
among three Bobs if they share singlets $|\Psi_{-}\rangle$, 
as is depicted in Fig.~2. 
Since this scheme is a sequence of partial teleportations 
we need to generally evaluate 
the fidelities of clones after every step of the procedure. 
To find $R_{1},\ldots,R_{M-1}$ for a symmetric distribution of $M$
clones we have to solve a set of quadratic equations for fidelities with condition 
$F_{1},\ldots,F_{M}$, which 
can be performed numerically. Analytically, we can present the simplest example for $M=3$, in which
we set $R_{1}=3/8$, $R_{2}=1/3$ to obtain symmetric distribution. 
At result, we obtain the same fidelity of all three clones $F=29/38\approx 0.763$. 
It is slightly worse in comparison with 
the fidelity $F=7/9\approx 0.777$ of optimal $1\rightarrow 3$ cloning \cite{Gisin97,Bruss98}. 
Thus we cannot generally use a sequence of 
asymmetric optimal cloners to distribute information to many users in optimal way.  
It is apparently dissimilar with classical-like universal 
cloning when we with a given probability 
swap an unknown state to one from the $M$ users and to the others we send completely randomized state. 
In this case, the fidelity of cloning $F_{M}=\frac{1}{2}\left(1+\frac{1}{M}\right)$ is always 
less than optimal universal cloning
$F_{1\rightarrow M}=\frac{2M+1}{3M}$ but this classical-like $1\rightarrow M$ 
cloning can be implemented by a sequence of $1\rightarrow 2$ classical-like cloners.  


\section{Optimal $N\rightarrow N+1$ cloning at a distance}

The setup for symmetric teleportation from $N$ identical replicas of input state $|\Psi\rangle_{S}$ 
on single distant copy is depicted in Fig.~3. It is an extension of the previous setup by 
additional unbalanced beam splitters $BS_2-BS_N$ placed in mode $I$ which have specific 
reflectivities $R_{2},\ldots,R_{N}$. Thus we implement the following sequence of projective measurements  
$\Pi^{-}_{SN,I}(R_{N})\ldots\Pi^{-}_{S2,I}(R_{2})\Pi^{-}_{S1,I}(R_{1})
\otimes 1_{S'}$
on a state of total system $|\Psi\rangle_{SN}\ldots|\Psi\rangle_{S2}|\Psi
\rangle_{S1}|\Psi_{-}\rangle_{IS'}$
and optimize the reflectivities $R_{n}$ in such a way to achieve symmetric 
distribution of state $|\Psi\rangle$ in $N+1$ copies. To obtain it we must adjust the reflectivities 
according to 
\begin{equation}\label{refl1}
R_{n}=\frac{1}{n+2},
\end{equation}
where $n=1,\ldots,N$ and then Alice obtains $N$ optimal clones 
of input state in modes $S_{1},\ldots,S_{N}$, 
single anti-clone in mode $I$ and on the other hand, Bob has at a distance a single clone 
in mode $S'$. To prove this, we calculate the  
probabilities that state $|\Psi_{\bot}\rangle$ can be detected
in the particular output modes $S'$ and $I$
\begin{equation}\label{1set}
p_{S'}^{\bot}=\frac{1}{P(N)}\prod_{k=1}^{N}(1-2R_{k})^2,\,\,
p_{I}^{\bot}=\frac{1}{P(N)}\prod_{k=1}^{N}(1-R_{k})^{2}. 
\end{equation}
and in modes $S_n$ 
\begin{equation}
p^{\bot}_{S_{n}}=\frac{1}{P(N)}R_{n}^{2}\prod_{k=1}^{n-1}(1-R_{k})^{2}\prod_{k=n+1}^{N}(1-2R_{k})^{2}.
\end{equation}
For $R_{k}$ given by (\ref{refl1}), the total probability $P(N)$ of success can be determined 
from normalization condition $\sum_{n=1}^{N}p^{\bot}_{S_{n}}+p_{S'}^{\bot}+p_{I}^{\bot}=1$
and is equal to $P(N)=4/((N+1)(N+2))$.
Consequently, the fidelity of $n$-th clone is $F_{n}=1-p^{\bot}_{S_{n}}$
and inserting reflectivities (\ref{refl1}) we can simply 
prove that all photons in the modes $S_1,\ldots,S_{N}$ 
have the same fidelity equal to (\ref{fid1}). The distant Bob's clone in the mode $S'$ has a fidelity 
$F_{n+1}^{S'}=1-p_{S'}^{\bot}$
and using (\ref{refl1}) we can subsequently prove that the clone has fidelity equal to (\ref{fid1}).  
Apart from  $N+1$ clones the setup produces also single anti-clone in the mode $I$. 
Using (\ref{1set}) for fidelity between the anti-clone and state $|\Psi_{\bot}\rangle$, 
we can simply calculate that the final anti-clone has the fidelity 
\begin{equation}
F_{I}=\frac{N+1}{N+2}.
\end{equation}
As the number of replicas increases we obtain a better and still optimal approximation 
of the U-NOT gate for $N$ replicas. For demonstration of local U-NOT with $N$ replicas 
we need no source of entanglement and 
only $N$ unbalanced beam splitters having reflectivity according to Eq.~(\ref{refl1}) 
with single port in completely random polarization state is required. 

\begin{figure}
\centerline{\psfig{width=8.0cm,angle=0,file=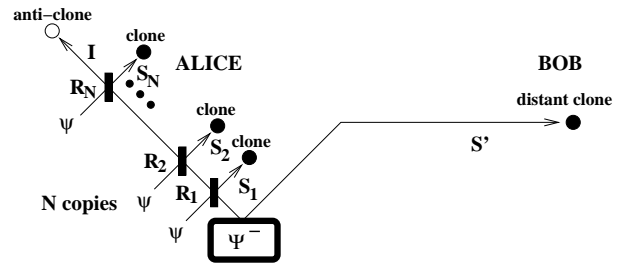}}
\caption{Scheme of conditional partial teleportation as $N\rightarrow N+1$ symmetric optimal cloning.}
\end{figure}

\section{Experimental implementations}

An experimental realization of partial teleportation as optimal asymmetric $1\rightarrow 2$ cloning of 
a polarization state of a photon depicted in Fig.~1 is straightforward modification of 
a well-known previous experiment demonstrating the total teleportation of polarization state 
\cite{teleexp}. We only have to be able to control the reflectivity of the beam 
splitter in the Bell-state measurement. 
Using single pair of photons in state $|\Psi_{-}\rangle$ and more input photons prepared as 
the identical replicas which can be directly extracted from pump beam by strong attenuation 
$|\Psi\rangle$ we can implement teleportation as $N\rightarrow N+1$ optimal cloning.
In this way, we also experimentally demonstrate the usefulness of the multiple copies 
to locally realize U-NOT gate with a higher fidelity. 
 
These schemes can be also implemented in the experiments on long-distance 
teleportation of time-bin qubit \cite{Marcikic03}. 
A time-bin qubit is a quantum superposition of a photon in a different time-bins
$|\Psi\rangle_{S}=\alpha|1,0\rangle_{S}+\beta|0,1\rangle_{S}$, 
where basis state $|1,0\rangle_{S}$ corresponds 
to first time-bin and $|0,1\rangle_{S}$ to the second one. 
To teleport time-bin qubit Alice and Bob use shared time-bin entangled state 
\begin{equation}\label{time_bin_ent}
|\Phi_{+}\rangle=\frac{1}{\sqrt{2}}(|1,0\rangle_{I}|1,0\rangle_{S'}+|0,1\rangle_{I}|0,1\rangle_{S'}),
\end{equation}
which can be produced from type I nonlinear down-conversion, 
where the pump pulse is splitted to two separate ones by unbalanced Michelson interferometer. 
If we restrict only to cases when two photons are 
emitted either by first pumping pulse or second one we have exactly state (\ref{time_bin_ent}).
The Bell-state measurement was performed by mixing of two time-bin qubits in balanced 
fiber coupler followed by two single photon detectors and 
if both the detectors register photons in different time-bins the teleportation 
(up to an unitary operation on Bob side) has been successfully performed \cite{Marcikic03}.

To implement our idea of partial teleportation to time-bin qubit 
we need only an optical fiber coupler with variable coupling for the Bell-state 
projection. If we take into account only detection events when both detectors register only single photon 
in different time-bins we can describe an action of the variable coupler on basis states 
$|1,0\rangle$ and $|0,1\rangle$ by the same relations as in 
Eqs.~(\ref{unBS}). A calculation of partial teleportation of time-bin qubit can be 
done with the help of previous analysis. 
Let us consider thought unitary operation $U_{S'}$ which converts the state (\ref{time_bin_ent}) to 
the state $\frac{1}{\sqrt{2}}(|1,0\rangle_{I}|0,1\rangle_{S'}-|0,1\rangle_{I}|1,0\rangle_{S'})$. 
This operation consists of mutual flip of basis states 
$|1,0\rangle_{S'}\leftrightarrow |0,1\rangle_{S'}$ and phase shift 
$|1,0\rangle_{S'}\rightarrow -|1,0\rangle_{S'}$, $|0,1\rangle_{S'}\rightarrow |0,1\rangle_{S'}$.
Then we obtain analogical teleportation scheme with shared $|\Psi^{-}\rangle$-like state as has 
been discussed above. 
After successful teleportation we implement second thought unitary operation $U^{\dag}_{S'}$ 
on time-bin qubit $S'$ and due to $U^{\dag}_{S'}U_{S'}=1$ we obtain in fact the same result 
as with the state (\ref{time_bin_ent}) shared between Alice and Bob. 
Thus after this teleportation we have the same fidelities $F_{S}(R)$ and $F_{I}(R)$ with 
input state $|\Psi\rangle_{S}$, 
however the Bob time-bin qubit is in the state having the $F_{S'}(R)$ with transformed state 
$|\Psi'\rangle=\alpha|0,1\rangle_{S'}-\beta|1,0\rangle_{S'}$. 
Therefore Bob has to perform unitary physical operation $U_{S'}$ on 
time-bin qubit $S'$ to obtain the demanded state having fidelity $F_{S'}(R)$ 
with state $|\Psi\rangle_{S}$. 

In this paper we propose two extended conditional teleportation schemes 
as asymmetric $1\rightarrow 2$ and $N\rightarrow N+1$ cloning at a distance 
which can be straightforwardly implemented in the recent quantum teleportation experiments.
Further, we discuss an experiment on the conditional 
LOCC reversibility of the partial teleportation and 
a conditional realization of optimal U-NOT operation on the multiple copies. 

\medskip
\noindent {\bf Acknowledgments}
The work was supported by the 
project LN00A015, Post-doc grant 202/03/D239 of 
Czech grant agency and CEZ: J14/98 of the Ministry of Education of Czech Republic. I would like to
thank to  Markus Aspelmeyer, \v Caslav Brukner, Nicolas Cerf, Jarom\' ir Fiur\' a\v sek, Petr Marek and 
L. Mi\v sta Jr., for stimulating and fruitful discussions.


\end{document}